\documentstyle[12pt]{article}
\newcommand{\resection}[1]{\setcounter{equation}{0}\section{#1}}

\def\MPL #1 #2 #3 {{\sl Mod.~Phys.~Lett.}~{\bf#1} (#3) #2}
\def\NPB #1 #2 #3 {{\sl Nucl.~Phys.}~{\bf B#1} (#3) #2}
\def\PLB #1 #2 #3 {{\sl Phys.~Lett.}~{\bf B#1} (#3) #2}
\def\PR #1 #2 #3 {{\sl Phys.~Rep.}~{\bf#1} (#3) #2}
\def\PRD #1 #2 #3 {{\sl Phys.~Rev.}~{\bf D#1} (#3) #2}
\def\PRL #1 #2 #3 {{\sl Phys.~Rev.~Lett.}~{\bf#1} (#3) #2}
\def\RMP #1 #2 #3 {{\sl Rev.~Mod.~Phys.}~{\bf#1} (#3) #2}
\def\ZPC #1 #2 #3 {{\sl Z.~Phys.}~{\bf C#1} (#3) #2}
\def\IJMP #1 #2 #3 {{\sl Int.~J.~Mod.~Phys.}~{\bf#1} (#3) #2}
\def\Zint{{Z \kern -.45 em Z}}
\def\complex{{\kern .1em {\raise .47ex \hbox
{$\scriptscriptstyle |$}}
\kern -.4em {\rm C}}}
\def\real{{\vrule height 1.6ex width 0.05em depth 0ex
\kern -0.06em {\rm R}}}

\newcommand{\be}{\begin{equation}}
\newcommand{\ee}{\end{equation}}
\newcommand{\bea}{\begin{eqnarray}}
\newcommand{\eea}{\end{eqnarray}}
\newcommand{\nn}{\nonumber}
\newcommand{\dd}{\displaystyle}

\newcommand{\AAA}{{\cal A}}

\def\esp#1{e^{#1}}

\def\bra#1{\langle #1 |}
\def\ket#1{| #1\rangle }
\begin{document}

\vspace*{.5cm}
\begin{center}
  \begin{Large}
  \begin{bf}
Projective Group Algebras\\
  \end{bf}
  \end{Large}
\end{center}
  \vspace{5mm}
\begin{center}
  \begin{large}
R. Casalbuoni\footnote{On leave from
Dipartimento di Fisica, Universit\`a di Firenze,
I-50125 Firenze, Italia}\\
  \end{large}
D\'epartement de
Physique Th\'eorique, Universit\'e de Gen\`eve\\ CH-1211 Gen\`eve
4, Suisse \\ {\tt{e-mail: CASALBUONI@FI.INFN.IT}}\\
\end{center}
\vspace{0.2cm}
\begin{center}
  \begin{bf}
  ABSTRACT
  \end{bf}
\end{center}
  \vspace{0.2cm}
\noindent
In this paper we apply a recently proposed algebraic theory of integration  
to projective group algebras. These structures have received some attention 
in connection with the compactification of the $M$ theory on noncommutative 
tori. This turns out to be an interesting field of applications, since the 
space $\hat G$ of the equivalence classes of the vector unitary irreducible 
representations of the group under examination becomes, in the projective   
case, a prototype of noncommuting spaces. For vector representations the 
algebraic integration is equivalent to integrate over $\hat G$. However, its 
very definition is related only at the structural properties of the group 
algebra, therefore it is well defined also in the projective case, where the 
space $\hat G$ has no classical meaning. This allows a generalization of the 
usual group harmonic analysis. A particular attention is given to abelian 
groups, which are the relevant ones in the compactification problem,  since 
it is possible, from the previous results, to establish a simple 
generalization of the ordinary calculus to the associated noncommutative 
spaces.

\vspace{.2 cm}
\begin{center}
UGVA-DPT 1998/04-1001
\end{center}
\vspace{.5 cm}
\newpage


\newpage
\resection{Introduction}

A very large class of algebras is obtained by considering
the projective representations of an arbitrary group $G$.
 These algebras have been recently
investigated in relation with the problem of
compactification within the $M$-theory \cite{tori1,tori2}.
In fact, to deal with this problem one needs to extend the
matrix valued coordinates of the $D0$ branes, $X^\mu_{i_1
i_2}$, $i_1,i_2=1,\cdots,N$, $\mu=0,1,\cdots,D$, where $D+1$
is the number of space-time dimensions, to matrices,
$X^\mu_{(i_1,a_1)(i_2,a_2)}$, where $a_1$ and $a_2$ are
elements of a group (in general discrete) $G$ of euclidean
motions in the subspace  $\real^n\in\real^D$,  $n<D$, to be
compactified \cite{compact}. The compactification condition
on the corresponding coordinates becomes a requirement of
symmetry of the theory under the unitary transformation
\be
U^{-1}(a)X^\mu U(a)= X^\mu+d^\mu_a
\ee
The quantity $d^\mu_a$ is the translation which defines  the
lattice structure of $\real^n$ induced by $G$. In this way,
the theory is effectively defined on the coset space
$\real^n/G$ (for instance, for $G=Z^n$, the
$n$-dimensional torus). The matrix $U(a)$ acts on the group
indices of $X_\mu$, and must satisfy the group
multiplication rule
\be
U(a)U(b)=\esp{i\alpha(a,b)}U(ab)
\ee
Therefore, the matrices $U(a)$ are in the  regular
projective representation of the group $G$, that is they
belong to the algebra of the right (or left)
multiplications. Besides the interest of projective group
algebras in  the previous problem, they are an excellent
ground for studying the nature of the algebraic theory of
integration (ATI) that we have recently considered in ref
\cite{integrale}. Strictly speaking a group algebra is
defined by taking formal linear combinations of the group
elements with coefficients in a given field $F$ (here we
will take $F=\complex$). A more convenient way to do this
construction, is to consider in place of the abstract group
$G$, a linear vector representation, $x_\lambda(a)$, $a\in
G$, where $\lambda$ specifies which representation we are
using. The linear combinations of $x_\lambda(a)$ span a
vector space $\AAA(G)_\lambda$, which has also an algebra
structure induced by the group product
\be
x_\lambda(a)x_\lambda(b)=x_\lambda(ab)
\label{0}
\ee
In these cases it is convenient to introduce the  set $\hat
G$ of the equivalence classes of the unitary irreducible
representations of $G$ (in the following, for simplicity, we
will consider only discrete and compact groups). From this
point of view, an arbitrary function on the algebra
\be
\hat f(\lambda)=\sum_{a\in G}f(a) x_\lambda(a)
\label{1}
\ee
can be regarded as follows:  the quantities $f(a)$ define a
mapping from $G$ to $\complex$ and eq. (\ref{1}) can be
thought to define the  Fourier transform of $f(g)$ with
respect to the representation $x_\lambda$. At each point
$\lambda\in \hat G$, $\hat f(\lambda)$ takes values in
$\AAA(G)_\lambda$. Therefore, the concept of group algebra
is strictly related to the harmonic analysis over a group.
From an algebraic point of view, the elements $x_\lambda(a)$
can be  seen as the generators of the algebra, since any
element of $\AAA(G)_\lambda$ can be expressed as a linear
combination of them. This scheme can be extended to the case
of projective representations, where the only difference is
the product rule (\ref{0}) which is modified by a phase
factor (since we are considering unitary representations)
\be
x_\lambda(a)x_\lambda(b)=\esp{i\alpha(a,b)}x_\lambda(ab)
\ee
$\alpha(a,b)$ is called a cocycle. In this case we will speak of
projective  group algebras. The structure of these algebras is
rather simple, but also very rich, and therefore it is possible
to use them as a laboratory to study the properties of the ATI.
In fact, we will show that this theory  allows us to invert eq.
(\ref{1}). This means that, in the vector case, the integration defined by 
the ATI is equivalent to "sum", or "integrate" over all the unitary
irreducible representations of $G$. In other words it defines an
integration measure in the space $\hat G$. This is particularly
simple in the case of the abelian groups, where $\hat G$ is also
an abelian group (the dual of $G$ in the Pontryagin sense
\cite{pont}). For instance, for $G=\real$, we have $\hat
G=\real$, and the integral defined  by the ATI becomes the
Lebesgue integral over $\hat G=\real$. However the situation is
completely different in the case of projective representations.
In fact, already for abelian groups, the space $\hat G$ becomes a
noncommutative space. By this we mean the following: a vector
representation of $G=\real^D$
\be
x_{\vec q}\;(\vec a)=\esp{-i\vec q\cdot\vec a},~~~\vec a\in \real^D,~~~\vec
q\in
\hat G=\real^D
\ee
can be extended to a projective  representation of
$\real^D$, at the price of considering $\vec q$ not as
numbers, but operators with non-vanishing commutation
relations
\be
[q_i.q_j]=i\eta_{ij}
\ee
where  $\eta_{ij}$ is related to the cocycle
which characterizes the projective representation. In this
way, the space $\hat G$ looses its classical features and
becomes an example of  noncommutative space. At the same
time the ordinary notion of integration over this space is
meaningless. On the other hand,  the ATI is based only on
the algebraic properties  of the projective group algebra
$\AAA(G)$ and not on the representation $x_\lambda$ chosen
to define it. In fact it depends only on the properties of
the regular representation. In the classical case the
integration through functions on $\hat G$ (the
representations $x_\lambda$) allows the reconstruction of
the integration over $\hat G$. In the noncommuting case the
definition remains valid also if the "space" $\hat G$ has no
meaning, since we are  using the  space $\AAA(G)$ for our
construction.

The paper is organized as follows. In Section 2  we
establish the integration properties for a projective group
algebra, following the prescriptions of the ATI. To this end
we need to determine a set of relations for the cocycles of
the algebra that are necessary to obtain the integration
rules. Section 3 is devoted to understand the meaning of the
ATI, by examining the special case of vector
representations. Here we prove that for discrete and compact
groups, the ATI corresponds to integrate over the space
$\hat G$, in such a way to provide the inversion formula for
the Fourier transformation on the group, and to ensure the
validity of the Plancherel formula. We discuss also how, in
the case of projective representations, the Fourier analysis
is generalized through the use of the ATI. The space $\hat
G$ looses its meaning, but the ATI is still valid, and
therefore it can be seen as a theory of integration over a
noncommutative space $\hat G$. In Section 4 we consider the
case of projective representations of abelian groups. In
particular we compare our previous Fourier analysis with the
more conventional one made in terms of the characters of the
abelian groups. In this case, the Moyal product among the
functions on the group $G$ arises in a very natural way. But
this is avoided in our formulation, since we use the projective
representations for the expansion rather than the
characters. We show also that it is possible to introduce a
derivation on the projective group algebra for abelian
groups, such that its integral (in the sense of the ATI) is
zero. This allows to extend the standard calculus to this
noncommutative case. Explicit formulas are given  for
$G=\real^D$,  $G=Z^D$, and $G=Z_n^D$.

\resection{Projective group algebras}

Let us start defining a projective  group algebra. We
consider an arbitrary projective linear representation,
$a\to x(a)$, $a\in G$, $x(a)\in \AAA(G)$, of a given
group $G$. The representation $\AAA(G)$ defines in a natural
way an associative algebra with identity  (it is closed
under multiplication and it defines a generally complex
vector space). This algebra will be
 denoted by $\AAA(G)$. The elements of the algebra
are given by the combinations
\be
\sum_{a\in G}f(a) x(a)
\label{groupalgebra}
\ee
For a group with an infinite number of elements, there is no
a unique definition of such an algebra. The one defined in
eq. (\ref{groupalgebra}) corresponds to consider a formal
linear combination of a finite number of elements of $G$.
This is very convenient because we will not be concerned
here with topological problems. Other definitions correspond
to take complex functions on $G$ such that
\be
\sum_{a\in G}|f(g)|<\infty
\ee
Or, in the case of compact groups, the sum is defined in
terms of the Haar
 invariant measure. When necessary we will
be more precise about this point. The basic product rule of the
algebra follows from the group property
\be
x(a)x(b)=\esp{i\alpha(a,b)}x(ab)
\label{algebra}
\ee
where $\alpha(a,b)$ is called a cocycle. This is
constrained, by the requirement of associativity of the
representation, to satisfy

\be
\alpha(a,b)+\alpha(ab,c)=\alpha(b,c)+\alpha(a,bc)
\label{cocycle1}
\ee
Changing the element $x(a)$ of the algebra by a phase
factor $\esp{i\phi(a)}$, that is, defining
\be
x'(a)=\esp{-i\phi(a)}x(a)
\ee
we get
\be
x'(a)x'(b)=\esp{i(\alpha(a,b)-\phi(ab)+\phi(a)+\phi(b))}x'(ab)
\ee
This is equivalent to change the cocycle to
\be
\alpha'(a,b)=\alpha(a,b)-[\phi(ab)-\phi(a)-\phi(b)]
\ee
In particular, if $\alpha(a,b)$ is of the form
$\phi(ab)-\phi(a)-\phi(b)$, it can be transformed to zero,
and therefore the corresponding projective representation is
isomorphic to a vector one. For this reason the combination
\be
\alpha(a,b)=\phi(ab)-\phi(a)-\phi(b)
\ee
is called a trivial cocycle. Let us now discuss some
properties of the cocycles. We start from the relation ($e$
is the identity element of $G$)
\be
x(e)x(e)=\esp{i\alpha(e,e)}x(e)
\ee
By the transformation $x'(e)=\esp{-i\alpha(e,e)}x(e)$, we
get
\be
x'(e)x'(e)=x'(e)
\ee
Therefore we can assume
\be
\alpha(e,e)=0
\ee
Then, from
\be
x(e)x(a)=\esp{i\alpha(e,a)}x(a)
\ee
multiplying by $x(e)$ to the left, we get
\be
x(e)x(a)=\esp{i\alpha(e,a)}x(e)x(a)
\ee
implying
\be
\alpha(e,a)=\alpha(a,e)=0
\ee
where the second relation is obtained in analogous way.
Now, taking $c=b^{-1}$ in eq. (\ref{cocycle1}), we get
\be
\alpha(a,b)+\alpha(ab,b^{-1})=\alpha(b,b^{-1})
\label{cocycle2}
\ee
Again, putting $a=b^{-1}$
\be
\alpha(b^{-1},b)=\alpha(b,b^{-1})
\ee
We can go farther by considering
\be
x(a)x(a^{-1})=\esp{i\alpha(a,a^{-1})}x(e)
\ee
and defining
\be
x'(a)=\esp{-i\alpha(a,a^{-1})/2}x(a)
\ee
from which
\be
x'(a)x'(a^{-1})=\esp{-i\alpha(a,a^{-1})}x(a)x(a^{-1})=x(e)=x'(e)
\ee
Therefore we can transform $\alpha(a,a^{-1})$ to zero
without changing the definition of $x(e)$,
\be
\alpha(a,a^{-1})=0
\label{2.21}
\ee
As a consequence,  equation (\ref{cocycle2}) becomes
\be
\alpha(a,b)+\alpha(ab,b^{-1})=0
\label{cocycle3}
\ee
We can get another relation using $x(a^{-1})=x(a)^{-1}$
\bea
 x(a^{-1})x(b^{-1})&=&\esp{i\alpha(a^{-1},b^{-1})}x(a^{-1}b^{-1})=
 x(a)^{-1}x(b)^{-1}\nn\\&=&(x(b)x(a))^{-1}=
 \esp{-i\alpha(b,a)}x(a^{-1}b^{-1})
\eea
from which
\be
\alpha(a^{-1},b^{-1})=-\alpha(b,a)
\ee
and together with eq. (\ref{cocycle3}) we get
\be
\alpha(ab,b^{-1})=\alpha(b^{-1},a^{-1})
\label{cocycle4}
\ee
This relation will be very useful in the following. From the
product rule
\be
x(a)x(b)=\esp{i\alpha(a,b)}x(ab)=\sum_{c\in G} f_{abc}x(c)
\ee
we get the structure constants of the algebra
\be
f_{abc}=\delta_{ab,c}\esp{i\alpha(a,b)}
\ee
The delta function is defined according to the nature of the sum over the
group elements.

In order to define an integration over $\AAA(G)$, according to the rules of 
the ATI, we start introducing a ket with elements given by $x(a)$, that is 
$\ket {x}_a=x(a)$, and the
corresponding transposed bra $\bra x$. Then, we define left
and right multiplications as
\be
R(a)\ket x=\ket x x(a),~~~~\bra x L(a)= x(a) \bra x
\ee
where we have used the shorthand notation
$R(x(a))=R(a)$, and similarly for $L(a)$. From the algebra product, we get
immediately
\be
(R(a))_{bc}=f_{bac}=\delta_{ba,c}\esp{i\alpha(b,a)},~~~~
(L(a))_{bc}=f_{acb}=\delta_{ac,b}\esp{i\alpha(a,c)}
\ee
A self-conjugated algebra is defined as one equipped with a matrix $C$ such
that \cite{algebra}
\be
C^T=C,~~~~CR(a)C^{-1}=L(a)
\label{2.30}
\ee
The second equation tells us that the bra $\bra{x C}$ is  an
eigenstate of $R(a)$
\be
\bra {x C}R(a)=x(a) \bra{x C}
\label{2.31}
\ee
with eigenvalue $x(a)$, and
\be
(xC)_a=x(b)C_{ba}
\ee
Therefore, the matrix $C$ can be determined by solving eq. (\ref{2.31}).
We get
\be
(xC)_b\delta_{ba,c}\esp{i\alpha(b,a)}=(xC)_{ca^{-1}}
\esp{i\alpha(ca^{-1},a)}=x(a)(xC)_c
\ee
By putting
\be
(xC)_a=k_ax(a^{-1})
\ee
we obtain
\be
k_{ca^{-1}}x(ac^{-1})\esp{i\alpha(ca^{-1},a)}=
k_c\esp{i\alpha(a,c^{-1})}x(ac^{-1})
\ee
Then, from eq. (\ref{cocycle4})
\be
k_{ca^{-1}}=k_c
\ee
Therefore $k_a=k_e$, and assuming $k_e=1$, it follows
\be
(xC)_a=x(a^{-1})=x(a)^{-1}
\ee
giving
\be
C_{a,b}=\delta_{ab,e}
\ee
This shows also that
\be
C^T=C
\ee
at least in the cases of discrete and compact groups.
The mapping $C:\,\AAA\to\AAA$ is an involution of the
algebra. In fact, by defining
\be
x(a)^*=x(b)C_{b,a}=x(a^{-1})=x(a)^{-1}
\label{involution}
\ee
we have $(x(a)^*)^*=x(a)$, and $x(b)^*x(a)^*=(x(a)x(b))^*$.
In refs. \cite{integrale,algebra}  we have shown that for a
self-conjugated algebra, it is possible to define an
integration rule in terms of the matrix $C$
\be
\int_{(x)}x(a)=C_{e,a}^{-1}=\delta_{e,a}
\label{ATI}
\ee
From this definition and the eq. (\ref{2.30}) it follows
\cite{integrale,algebra}
\be
\int_{(x)}\ket x\bra {xC}= \int_{(x)}\ket {xC}\bra {x}=1
\ee
Therefore we are allowed to expand a function on the group
($\ket f_a=f(a)$) as
\be
f(a)=\int_{(x)}x(a^{-1})\langle x|f\rangle
\label{2.43}
\ee
with $\langle x|f\rangle=\sum_{b\in G} x(b)f(b)$. It is also
possible to define a scalar product among functions on the
group. Defining,  $\bra f_a=\bar f(a)$, where $\bar f(a)$ is the
complex conjugated of $f(a)$, we put
\be
\langle f |g\rangle=\int_{(x)} \langle f|xC\rangle\langle x|g\rangle=
\sum_{a\in G} f^*(a) g(a)
\ee
with
\be
f^*(a)=\langle f|xC\rangle=\bar f(a^{-1})
\ee
It is important to stress that this definition depends only on
the algebraic properties of $\AAA(G)$ and not on the specific
representation chosen for this construction.

\resection{What is the meaning of the algebraic integration?}

As we have said in the previous Section, the integration formula we have
obtained  is independent on the group representation we started with. In
fact, it
is based only on the structure of right and left multiplications, that is on
the abstract algebraic product. This independence on the representation
suggests that in some way we are "summing" over all the representations.
To understand this point, we will study in this Section vector
representations. To do that, let us introduce a label $\lambda$ for the
vector representation we are actually using to define $\AAA(G)$. Then a
generic function on $\AAA(G)_\lambda$
\be
\hat f(\lambda)=\sum_{a\in G} f(a)x_\lambda(a)
\label{F-transform}
\ee
can be thought as the Fourier transform of the function $f:
G\to\complex$. Using the algebraic integration we can invert this
expression (see eq. (\ref{2.43}))
\be
f(a)=\int_{(x_\lambda)}\hat f(\lambda) x_\lambda(a^{-1})
\label{inversion1}
\ee
But it is a well known result of the  harmonic analysis over the
groups, that in many cases it is possible to invert the Fourier
transform, by an appropriate sum over the representations. This
is true in particular for finite and compact groups. Therefore
the algebraic integration should be the same thing as summing or
integrating over the labels $\lambda$ specifying the
representation. In order to show that this is the case, let us
recall a few facts about the Fourier transform over the groups
\cite{pont}. First of all, given the group $G$, one defines the
set $\hat G$ of the equivalence classes of the irreducible
representations of $G$. Then, at each point $\lambda$ in $\hat G$
we choose a unitary representation $x_\lambda$ belonging to the
class $\lambda$, and define the Fourier transform of the function
$f:G\to
\complex$, by the eq. (\ref{F-transform}). In the case
of compact groups, instead of the sum over the group element
one has to integrate over the group by means of the
invariant Haar measure. For finite groups, the inversion
formula is given by
\be
f(a)=\frac 1{n_G}\sum_{\lambda\in\hat G} d_\lambda tr[\hat
f(\lambda)x_\lambda(a^{-1})]
\label{3.3}
\ee
where $n_G$ is the order of the group and $d_\lambda$ the
dimension of the representation $\lambda$. Therefore, we get
the identification
\be
\int_{(x)}\{\cdots\}=\frac 1{n_G}\sum_{\lambda\in\hat G}
d_\lambda tr[\{\cdots\}]
\label{correspondence}
\ee
A more interesting way of deriving this relation, is to take
in (\ref{F-transform}), $f(a)=\delta_{e,a}$, obtaining for
its Fourier transform,
$\hat\delta=x_\lambda(e)=1_{\lambda}$, where the last symbol
means the identity in the representation $\lambda$. By
inserting this result into (\ref{3.3}) we get the identity
\be
\delta_{e,a}=\frac 1{n_G}\sum_{\lambda\in\hat G} d_\lambda tr[\hat
x_\lambda(a^{-1})]
\ee
which, compared with eq. (\ref{ATI}), gives
(\ref{correspondence}). This  shows explicitly that the
algebraic integration for vector representations of $G$ is
nothing but the sum over the representations of $G$.

An analogous relation is obtained in the case of compact groups.
This can also be obtained by a limiting procedure from  finite
groups, if we insert $ 1/{n_G}$, the volume of the group, in the
definition of the Fourier transform. That is one defines
\be
\hat f(\lambda)=\frac 1 {n_G}\sum_{a\in G} f(a) x_\lambda(a)
\ee
from which
\be
f(a)=\sum_{\lambda\in\hat G} d_\lambda tr[\hat
f(\lambda)x_\lambda(a^{-1})]
\label{inversion2}
\ee
Then one can go to the limit by substituting the  sum over
the group elements
 with the  Haar measure
\be
\hat f(\lambda)=\int_G d\mu(a) f(a) x_\lambda(a)
\ee
The inversion formula (\ref{inversion2}) remains unchanged.
We see that in these cases the algebraic integration sums
over the elements of the space $\hat G$, and therefore it
can be thought as the dual of the sum over the group elements
(or the Haar integration for compact groups). By using the
Fourier transform (\ref{F-transform}) and its inversion
(\ref{inversion1}), one can easily establish the Plancherel
fromula. In fact by multiplying together two Fourier
transforms, one gets
\be
\hat f_1(\lambda)\hat f_2(\lambda)=
\sum_{a\in G}\left(\sum_{b\in G}f_1(b)f_2(b^{-1}a)\right)x_\lambda(a)
\label{convolution}
\ee
from which
\be
\int_{(x)}\hat f_1(\lambda)\hat f_2(\lambda)x_\lambda(a^{-1})=
\sum_{b\in G}f_1(b)f_2(b^{-1}a)
\ee
and taking $a=e$ we obtain
\be
\int_{(x)}\hat f_1(\lambda)\hat f_2(\lambda)=
\sum_{b\in G}f_1(b)f_2(b^{-1})
\label{pre-Plancherel}
\ee
This formula can be further specialized, by taking
$f_2\equiv f$ and for $f_1$ the involuted of $f$. That is
\be
\hat f^*(\lambda)=\sum_{a\in G} {\bar f}(a)x_\lambda(a^{-1})
\ee
where use has been made of eq. (\ref{involution}). Then,
from eq. (\ref{pre-Plancherel}) we get the Plancherel
formula
\be
\int_{(x)}\hat f^*(\lambda)\hat f(\lambda)=\sum_{a\in G}{\bar f}(a)f(a)
\label{Plancherel}
\ee
Let us also notice that eq. (\ref{convolution})  says  that
the Fourier transform of the convolution of two functions on
the group is the product of the Fourier transforms.

We will consider now  projective representations.  In this
case, the product of two Fourier transforms is given by
\be
\hat f_1(\lambda)\hat f_2(\lambda)= \sum_{a\in G} h(a) x_\lambda(a)
\label{product}
\ee
with
\be
h(a)=\sum_{b\in G} f_1(b)f_2(b^{-1}a)\esp{i\alpha(b,b^{-1}a)}
\label{convolution2}
\ee
Therefore, for projective representations, the convolution
product is  deformed due to the presence of the phase factor.
However, the Plancherel formula still holds. In fact, since in
\be
h(e)=\sum_{b\in G} f_1(b)f_2(b^{-1})
\ee
using eq. (\ref{2.21}), the phase factor disappears,
the previous derivation from eq. (\ref{pre-Plancherel}) to eq.
(\ref{Plancherel}) is still valid. Notice that eq.
(\ref{product}) tells us that the Fourier  transform of the
deformed convolution product of two functions on the group,
is equal to the product of the Fourier transforms.

\resection{The case of abelian groups}

In this Section we consider the case of abelian groups, and
we compare the Fourier analysis made in the framework of the
ATI with the more conventional one made in terms of the
characters. A fundamental property of the abelian groups is
that the set $\hat G$ of their vector unitary irreducible
representations (VUIR), is itself an abelian group, the dual
of $G$ (in the sense of Pontryagin \cite{pont}). Since the
VUIR's are one-dimensional, they are given by the characters
of the group. We will denote the characters of $G$ by
$\chi_\lambda(a)$, where $a\in G$, and $\lambda$ denotes the
representation of $G$. For what we said before, the
parameters $\lambda$ can be thought as the elements of the
dual group. The parameterization of the group element $a$
and of the representation label $\lambda$ are given in Table
1, for the most important abelian groups and for their dual
groups, where we have used the notation $a=\vec a$ and
$\lambda=\vec q$.

\begin{table}[t]
\begin{center}
\begin{tabular}{|c||c|c|c|c|}
\hline\hline
& & & &\\
 & $G=R^D$& $G=Z^D$& $G=T^D$& $G=Z_N^D$\\
 &$\hat G= R^D$&$\hat G= T^D$&$\hat G= Z^D$&$\hat G= Z_N^D$\\
 &&&&\\
 \hline\hline
 &&&&\\
 $\vec a$ & $-\infty\le a_i\le +\infty$ & $a_i=
 {\dd{\frac{2\pi m_i}L}}$ &
 $0\le a_i\le L $& $a_i= k_i,$\\
 && $m_i\in Z$ & & $0\le k_i\le n-1$\\
&&&&\\
\hline
 &&&&\\
 $\vec q$ & $-\infty\le q_i\le +\infty$ & $0\le q_i\le L $ &
 $q_i={\dd{\frac{2\pi m_i}L}}$&
  $q_i= {\dd{\frac{2\pi\ell_i}N}}$\\
 && &$m_i\in Z$  & $0\le \ell_i\le n-1$\\
&&&&\\
\hline\hline
\end{tabular}
\end{center}
\begin{center}
 Table 1: {\it Parameterization of the abelian group $G$ and of its dual
$\hat G$,
 for $G=\real^D$, $Z^D$, $T^D$, $Z_n^D$}.
\end{center}
\end{table}

The characters are given by
\be
\chi_{\lambda}(a)\equiv\chi_{\vec q}\;(\vec a)=\esp{-i\vec q\cdot\vec
a}
\ee
and satisfy the relation (here we  use the additive notation
for the group operation)
\be
\chi_\lambda(a+b)=\chi_\lambda(a)\chi_\lambda(b)
\label{4.3}
\ee
and the dual
\be
\chi_{\lambda_1+\lambda_2}(a)=\chi_{\lambda_1}(a)\chi_{\lambda_2}(a)
\ee
That is they define vector representations of the abelian
group $G$ and of its dual, $\hat G$. Also
we can easily check that the operators
\be
 D_{\vec q} \chi_{\vec q}\;(\vec a)=-i\vec a \chi_{\vec q}\;(\vec a)
 \label{4.4}
\ee
are derivations on the algebra (\ref{4.3}) of the characters
for any $G$ in Table 1.

We can use the characters to define the Fourier transform of
the function $f(g): G\to\complex$
\be
\tilde f(\lambda)=\sum_{a\in G} f(a)\chi_\lambda(a)
\ee
If we evaluate the Fourier transform of the deformed
convolution of eq. (\ref{convolution2}), we get
\be
\tilde h(\lambda)=\sum_{a\in G} h(a)\chi_\lambda(a)=
\sum_{a,b\in G}f(a)\chi_\lambda(a)
\esp{i\alpha (a,b)}g(b)\chi_\lambda(b)
\ee
In the case of vector representations the Fourier transform
of the convolution is the product of the Fourier transforms.
In the case of projective representations, the result, using
the derivation introduced before, can be written in terms of
the Moyal product (we omit here the vector signs)
\be
\tilde h(\lambda)=\tilde f(\lambda)\star\tilde
g(\lambda)=\esp{-i\alpha(D_{\lambda'},D_{\lambda''})}
\tilde f(\lambda')\tilde g(\lambda'')\Big|_{\lambda'=\lambda''=
\lambda}
\ee
Therefore, the Moyal product arises in a very natural way
from the projective group algebra. On the other hand,  we
have shown in the previous Section, that the use  of the
Fourier analysis in terms of the projective representations
avoids the Moyal product. The projective representations of
abelian groups  allow a derivation on the algebra, analogous
to the one in eq. (\ref{4.4}), with very special features.
In fact we check easily that
\be
\vec D x_\lambda(\vec a)=-i\vec a x_\lambda(\vec a)
\label{derivation}
\ee
is a derivation, and furthermore
\be
\int_{(x_\lambda)} \vec D x_\lambda(\vec a)=0
\ee
From this it follows, by linearity, that the integral of  $\vec D$  applied
to any function on the algebra is zero
\be
\int_{(x_\lambda)} \vec D \left(\sum_{a\in G}f(\vec a)
x_\lambda(\vec a)\right)=0
\label{byparts}
\ee
This relation is very important because, as  we have shown
in \cite{algebra},  the automorphisms generated by $\vec D$,
that is $\exp(\vec\alpha\cdot\vec D)$, leave invariant the
integration measure of the ATI (see also later on). Notice
that this derivation generalizes the derivative with respect
to the parameter $\vec q$, although this has no meaning in
the present case. In the case of nonabelian groups, a
derivation sharing the previous properties can be defined
only if there exists a mapping $\sigma: G\to
\complex$, such that
\be
\sigma(ab)=\sigma(a)+\sigma(b), ~~~~a,b\in G
\ee
since in this case, defining
\be
Dx(a)=\sigma(a) x(a)
\ee
we get
\bea
D(x(a)x(b))&=&\sigma(ab) x(a)x(b)=(\sigma(a)+\sigma(b))x(a)x(b)\nn\\&=&
(Dx(a))x(b)+x(a)(Dx(b))
\eea

Having
defined derivations and integrals one has all the elements for the harmonic
analysis on the projective representations of an abelian group.

Let us start considering $G=R^D$. In the case of vector representations we
have
\be
x_{\vec q}\;(\vec a)=\esp{-i\vec q\cdot\vec a}
\label{a0}
\ee
with $\vec a\in G$, and $\vec q\in\hat G=R^D$ labels the representation. The
Fourier transform is
\be
\hat f(\vec q)=\int d^D\vec a f(\vec a)\esp{-i\vec q\cdot\vec a}
\label{a1}
\ee
Here the Haar measure for $G$ coincides with the ordinary Lebesgue measure.
Also, since $\hat G=R^D$, we can invert the Fourier transform by using the
Haar measure on the dual group, that is, again the Lebesgue measure. In the
projective case, eq. (\ref{a0}) still holds true, if we assume $\vec q$ 
 as
a vector operator satisfying the commutation relations
\be
[q_i,q_j]=i\eta_{ij}
\label{CR}
\ee
with $\eta_{ij}$ numbers which can be related to the cocycle, by using
the Baker-Campbell-Hausdorff formula
\be
\esp{-i\vec q\cdot\vec a}\esp{-i\vec q\cdot\vec b}=
\esp{-i\eta_{ij}a_ib_j/2}\esp{-i\vec q\cdot(\vec a+\vec b)}
\ee
giving
\be
\alpha(\vec a,\vec b)=-\frac 1 2 \eta_{ij} a_i b_j
\ee
The inversion of the Fourier transform can now be obtained by the ATI in the
form
\be
f(\vec a)=\int_{(\vec q)}\hat f(\vec q)x_{\vec q}\;(-\vec a)
\ee
where the dependence on the representation is expressed in terms of $\vec q$,
thought now they are not coordinates on $\hat G$.
We recall that in this case, eq. (\ref{ATI}) gives
\be
\int_{(\vec q)}x_{\vec q}\;(\vec a)=\delta^D(\vec a)
\label{a2}
\ee
Therefore, the relation between the integral in ATI and the
Lebesgue integral in $\hat G$, in the vector case is
\be
\int_{(\vec q)}=\int\frac{d^D\vec q}{(2\pi)^D}
\ee
In the projective case the  right hand side of this relation has
no meaning, whereas the left hand side is still well defined.
Also, we cannot maintain the interpretation of  the $q_i$'s as
coordinates on the dual space $\hat G$. However, we can define
elements of $\AAA(G)$ having the properties of the $q_i$'s (in
particular satisfying eq. (\ref{CR})), by using the Fourier
analysis. That is we define
\be
q_i=\int d^D\vec a\left(-i\frac{\partial}{\partial a_i}\delta^D(\vec
a)\right)x_{\vec q}\;(\vec a)
\label{q-def}
\ee
which is  an element of $\AAA(G)$ obtained by Fourier
transforming a distribution  over $G$, which is a honestly
defined space. From this definition we can easily evaluate the
product
\be
q_i x_{\vec q}\;(\vec a)=\int d^D\vec b\left(-i\frac{\partial}{\partial
b_i}\delta^D(\vec
b)\right)x_{\vec q}\;(\vec b)x_{\vec q}\;(\vec a)
\ee
Using the algebra  and integrating by parts, one gets
the result
\be
q_ix_{\vec q}\;(\vec a)=i\nabla_i x_{\vec q}\;(\vec a)
\label{5.28}
\ee
where
\be
\nabla_i=\frac{\partial}{\partial a_i} +i\alpha_{ij} a_j
\ee
where $\alpha_{ij}=\alpha({\vec e}_{(i)},{\vec e}_{(j)})$,
with ${\vec e}_{(i)}$ an orthonormal basis in $\real^D$. In
a completely analogous way one finds
\be
x_{\vec q}\;(\vec a)q_i=i{\overline \nabla}_i x_{\vec q}\;(\vec a)
\ee
where
\be
{\overline\nabla}_i=\frac{\partial}{\partial a_i} -i\alpha_{ij} a_j
\ee
Then, we  evaluate the  commutator
\be
[q_i,\hat f(\vec q)]=\int d^D\vec a
\left[-i\left({\overline\nabla}_i-\nabla_i\right)f(\vec a)\right]x_{\vec
q}(\vec
a)
\ee
where we have done an integration by parts. We get
\be
[q_i,\hat f(\vec q)]=-2i\alpha_{ij}D_{q_j}\hat f(\vec q)
\label{5.30}
\ee
where $D_{q_j}$ is the derivation (\ref{derivation}), with
$q_j$ a reminder for the direction along wich the derivation
acts upon. In particular, from
\be
D_{q_j}q_i=\int d^D\vec a\left(-i\frac{\partial}{\partial a_i}\delta^D(\vec
a)\right)(-ia_j)\esp{-i\vec q\cdot \vec a}=\delta_{ij}
\ee
we get
\be
[q_i,q_j]=-2i\alpha_{ij}
\ee
in agreement with eq. (\ref{CR}), after the identification
$\alpha_{ij}=-\eta_{ij}/2$.

The automorphisms induced by the derivations
(\ref{derivation}) are easily evaluated
\be
S(\vec \alpha)x_{\vec q}\;(\vec a)=\esp{\vec\alpha\cdot D_{\vec q}}
x_{\vec q}\;(\vec a)=\esp{-i\vec\alpha\cdot\vec a}x_{\vec q}\;(\vec
a)=x_{\vec
q+\vec\alpha}(\vec a)
\label{5.23}
\ee
where the last equality follows  from
\be
\int d^D\vec a\left(-i\frac{\partial}{\partial a_i}\delta^D(\vec
a)\right)\esp{\vec\alpha\cdot D_{\vec q}}x_{\vec q}\;(\vec a)=
q_i+\alpha_i
\ee
Meaning that in the vector case,  $S(\vec \alpha)$ induces
translations in $\hat G$. Since $D_{\vec q}$ satisfies the
eq. (\ref{byparts}), it follows from \cite{algebra} that the
automorphism $S(\vec\alpha)$ leaves invariant the algebraic
integration measure
\be
\int_{(\vec q)}=\int_{(\vec q+\vec\alpha)}
\label{invariance}
\ee
This shows that it is possible to construct a calculus completely
analogous to the one that we have on $\hat G$ in the vector case,
just using the Fourier analysis following by the algebraic
definition of the integral. We can push this analysis a little
bit further by looking  at the following expression
\be
\int_{(\vec q)}\hat f(\vec q)\;q_i\;x_{\vec q}\;(-\vec a)=-i
\left(\frac{\partial}{\partial a_i}+i \alpha_{ij}a_j\right)f(\vec
a)
\label{magnetic}
\ee
where we have used  eq. (\ref{5.28}). In the case $D=2$ this
equation has a physical interpretation in terms of a  particle of
charge $e$, in a constant magnetic field $B$. In fact, the
commutators among canonical momenta are
\be
[\pi_i,\pi_j]=ieB\epsilon_{ij}
\ee
where $\epsilon_{ij}$ is the 2-dimensional Ricci tensor. Therefore,
identifying $\pi_i$ with  $q_i$, we get $\alpha_{ij}=-eB\epsilon_{ij}/2$.
The corresponding vector potential is given by
\be
A_i(\vec a)=-\frac 1 2\epsilon_{ij}Ba_j=\frac 1 {e}\alpha_{ij}a_j
\ee
Then, eq. (\ref{magnetic}) tells us that the operation $\hat
f(\vec q)\to \hat f(\vec q)q_i$, corresponds to  take the
covariant derivative
\be
-i\frac{\partial}{\partial a_i}+eA_i(\vec a)
\ee
of the inverse Fourier transform of $\hat f(\vec q)$.  An interesting remark 
is
that a translation in $\vec q$ generated by
$\exp(\vec\alpha\cdot \vec D)$, gives rise to a phase
transformation on  $f(\vec a)$. First of all, by using the invariance of the 
integration measure we can check that
\be
\hat f(\vec q+\vec\alpha)=\esp{\vec\alpha\cdot\vec D} \hat f(\vec q)
\label{equality}
\ee
In fact
\be
\int_{(\vec q)}\hat f(\vec q+\vec\alpha)x_{\vec q}\;(-\vec a)=
\int_{(\vec q-\vec\alpha)}\hat f(\vec q)x_{\vec q-\vec\alpha}\;(-\vec a)=
\esp{-i\vec\alpha\cdot\vec a}f(\vec a)
\ee
Then, we have
\be
\int_{(\vec q)} \left(\esp{\vec\alpha\cdot\vec D}\hat f(\vec q)\right)x_{\vec 
q}\;(-\vec a)=
\int_{(\vec q)} \hat f(\vec q)\left(\esp{-\vec\alpha\cdot\vec D} x_{\vec 
q}(-\vec a)\right)=
\esp{-i\vec \alpha\cdot\vec a}f(\vec a)
\ee
where we have made use of eq. (\ref{5.23}). This shows eq. (\ref{equality}),
and at the same time our assertion.  From eq.
(\ref{magnetic}), this is equivalent to a  gauge
transformation on the gauge potential ${\cal
A}_i=\alpha_{ij}a_j$, ${\cal A}_i\to{\cal A}_i-
\partial_i\Lambda$, with $\Lambda=\vec \alpha\cdot\vec a$.
Therefore, we see here explicitly the content of a
projective representation in the basis of the functions on
the group. One starts assigning the two-form $\alpha_{ij}$.
Given that, one makes a choice for the vector potential. For
instance in the previous analysis we have chosen
$\alpha_{ij}a_j$. Any possible projective representation
corresponds to a different choice of the gauge.  In the dual
Fourier basis  this corresponds to assign a fixed set of
operators $q_i$, with commutation relations determined by
the two-form. All the possible projective representations
are obtained by translating the operators $q_i$'s. Of
course, this is equivalent to say that the projective
representations are the central extension of the vector
ones, and that they are determined by the cocycles. But the
previous analysis shows that the projective representations
generate noncommutative spaces, and that the algebraic
integration, allowing us to define a Fourier analysis, gives
the possibility of establishing the calculus rules.

Consider now the case $G=Z^D$. Let  us introduce an
orthonormal basis on the square lattice defined by $Z^D$,
${\vec e}_{(i)}$, $i=1,\cdots,D$. Then, any element of the
algebra can be reconstructed in terms of a product of the
elements
\be
U_i=x({\vec e}_{(i)})
\label{definition}
\ee
corresponding to a translation along the  direction $i$ by
one lattice site. In general we will have
\be
x(\vec m)=\esp{i\theta(\vec m)}U_1^{m_1}\cdots U_D^{m_D}
\label{4.42}
\ee
with $\theta$ a convenient phase factor such to reproduce
correctly the phase in the algebra product. The quantities
$U_i$ play the same role of $\vec q$ of the previous
example. The Fourier transform is defined by
\be
\hat f(\vec U)=\sum_{\vec m\in Z^D} f(\vec m) x_{\vec U}(\vec m)
\ee
where the dependence on the representation is expressed in terms of
$\vec U$, denoting the collections of the $U_i$'s.
The inverse Fourier transform is defined by
\be
f(\vec m)=\int_{\vec U} \hat f(\vec U)x_{\vec U}(-\vec m)
\ee
where the integration rule is
\be
\int_{(\vec U)} x_{\vec U}(\vec m)=\delta_{\vec m,\vec 0}
\ee
Therefore, the Fourier transform of $U_i$ is simply $\delta_{\vec m, \vec
e_{(i)}}$.
The algebraic integration for the vector case is
\be
\int_{(\vec U)}\to \int_0^{L} \frac{d^D\vec q}{L^D}
\ee
Since the set $\vec U$ is within the generators of the
algebra, to establish the rules of the calculus is a very
simple matter. Eq. (\ref{definition}) is the definition of
the set $\vec U$, analogous to eq. (\ref{q-def}). In place
of eq. (\ref{5.30}) we get
\be
U_i\hat f(\vec U)U_i^{-1}= \esp{-2\alpha_{ij}D_j}\hat f(\vec
U)
\ee
Here $D_j$ is the $j$-th  component of the derivation $\vec
D$ which acts upon $U_i$ as
\be
D_iU_j=-i\delta_{ij}U_j
\ee
By choosing $\hat f(\vec U)=U_k$ we have
\be
U_iU_kU_i^{-1}U_k^{-1}=\esp{2 i\alpha_{ik}}
\ee
which is the analogue of  the commutator among the $q_i$'s.
The automorphisms generated by $\vec D$ are
\be
S(\vec\phi)x_{\vec U}(\vec m)=\esp{\vec\phi\cdot\vec
D}x_{\vec U}(\vec m)=
\esp{-i\vec\phi\cdot\vec m}x_{\vec U}(\vec m)
\ee
From which we see that
\be
U_i\to S(\vec\phi)U_i=\esp{-i\phi_i}U_i
\ee
This transformation corresponds to  a trivial cocycle.  As
in the case $G=\real^D$ it gives rise to a phase
transformation on the group functions
\be
\int_{(\vec U)}\left(\esp{\vec\alpha\cdot\vec D}\hat f(\vec U)\right)x_{\vec 
U}(-\vec m)=
\int_{\vec U)}\hat f(\vec U)\left(\esp{-\vec\alpha\cdot\vec D}x_{\vec U}
(-\vec m)\right)=
\esp{i\vec\phi\cdot\vec m} f(\vec m)
\ee
Of course, all these relations could be obtained by putting
$U_i=\exp(-iq_i)$, with $q_i$ defined as in the case $G=R^D$.

Finally, in the case $G=Z_n^D$, the situation is very much alike $Z^D$, that
is the algebra can be reconstructed in terms of a product of elements
\be
U_i=x({\vec e}_{\;(i)})
\ee
satisfying
\be
U_i^n=1
\ee
Therefore we will not repeat the previous analysis but we will consider only
the case $D=2$, where $U_1$ and $U_2$ can be expressed as \cite{hoppe}
\be
(U_1)_{a,b}=\delta_{a,b-1}+\delta_{a,n}\delta_{b,1},~~~~
(U_2)_{a,b}=\esp{\frac {2\pi i} n(a-1)}\delta_{a,b},~~~~a,b=1,\cdots,n
\ee
The elements of the algebra are reconstructed as
\be
x_{\vec U}(\vec m)=\esp{i\frac{\pi} n m_1m_2}U_1^{m_1}U_2^{m_2}
\ee
The cocycle is now
\be
\alpha(\vec m_1,\vec m_2)=-\frac{2\pi} n\epsilon_{ij} m_{1i}m_{2j}
\ee
In this case we can compare the algebraic integration rule
\be
\int_{\vec U} x_{\vec U}(\vec m)=\delta_{\vec m,\vec 0}
\ee
with
\be
Tr[x_{\vec U}(\vec m)]=n\delta_{\vec m,\vec 0}
\ee
A generic element of the algebra is a $n\times n$ matrix
\be
A=\sum_{m_1,m_2=0}^{n-1}c_{m_1m_2}x_{\vec U}(\vec m)
\ee
and therefore
\be
\int_{\vec U} A=\frac 1 n Tr[A]
\ee
In ref. \cite{algebra} we have shown that the algebraic
integration over the algebra of the $n\times n$ matrices
$\AAA_n$ is given by
\be
\int_{\AAA_n} A=Tr[A]
\ee
implying
\be
\int_{\vec U}A=\frac 1 n \int_{\AAA_n}A
\ee

\resection{Conclusions}

During the last few years many  theoretical hints came about the
possible relevance of noncommutative spaces in physics.   Among
them we recall that $D0$ branes in $M$ theory are described by
noncommuting $N\times N$ hermitian matrices. Within the same
framework, noncommutative spaces arise from compactification in
various dimensions. The noncommutative compactification has to do
with the use of projective representations of abelian groups of
motion acting on the subspace to be compactified. To study the
related geometry it would be important to dispose of tools
allowing to mimic the usual calculus in classical spaces. In this
paper we have done an attempt along this direction by using, as a
principal instrument, an algebraic theory of integration that we
have developed in some recent paper. Here we have shown that the
ATI allows to generalize the usual harmonic analysis over vector
group representations to the projective case. This is done in a
way which is very much alike the spirit of noncommutative geometry 
\cite{connes}. That is,
the ATI deals with the algebra of functions, in this case with
the algebra of the projective representations, rather than with
the base space, which is the noncommutative
analogue of the space of the equivalence classes of the
representations of the group under consideration. For
abelian groups we have shown that it is possible, through the
harmonic analysis based on the ATI, to extend the usual tools of
the calculus over classical spaces, as derivatives, integrals,
and so on. For instance, it is possible to give a meaning to the
analogue of concepts as space coordinates, in terms of
distributions over the group.
Many of the concepts used in this study  need a more refined
mathematical analysis, but we feel that these ideas may have some interest 
for physical applications of noncommutative geometry.

\medskip
\begin{center}
{\bf Acknowledgements}
\end{center}
\medskip
The author would like  to thank
 Prof. J. P. Eckmann, Director of the Department of Theoretical
  Physics of the University
of Geneva, for the very kind hospitality.


\newpage
\begin{thebibliography}{99}





\bibitem{tori1}
A. Connes, M.R. Douglas and A. Schwarz, {\tt hep-th/9711162};
P.-M. Ho, Y.-Y. Wu and Y.-S. Wu, {\tt hep-th/9712201};
P.-M. Ho and Y.-S. Wu, {\tt hep-th/9801147}.


\bibitem{tori2}
R. Casalbuoni, Univ. of Geneva preprint, UGVA-DPT 1998/01-996, {\tt
hep-th/9801170}.


\bibitem{compact}
E.G. Gimon and J. Polchinski, Phys. Rev. {\bf D54} (1996) 1667;
C.V. Johnson and R.C. Myers, {\tt hep-th/9610140};
M.R. Douglas and G. Moore, {\tt hep-th/9603167};
W. Taylor IV, Phys. Lett. {\bf B394} (1997) 283.

\bibitem{integrale}
R.Casalbuoni, Int. J. Mod. Phys. {\bf A12} (1997) 5803,
{\tt physics/9702019}.


\bibitem{pont}
A.A. Kirillov, {\it \'El\'ements de la Th\'eorie des Repr\'esentations},
\'Editions MIR,  Moscou (1974); {\it ibidem} {\it Representation Theory and
Noncommutative Harmonic Analysis I}. Springer-Verlag (1994).

\bibitem{algebra}

R. Casalbuoni, Univ. of Geneva preprint, UGVA-DPT 1998/03-1000, {\tt
physics/9803024}

\bibitem{hoppe}
B. de Wit, J. Hoppe and H. Nicolai, Nucl. Phys. {\bf B305} (1988) 545;
D. Fairlie, P. Fletcher and C. Zachos, J. Math. Phys. {\bf 31} (1990)
1088; J. Hoppe, Int. J. Mod. Phys. {\bf A4} (1989) 5235.

\bibitem{connes}
A. Connes, {\it  Noncommutative geometry},
Academic Press (1994).


\end{thebibliography}
\end{document}